\documentclass[pra, twocolumn,showpacs,superscriptaddress,groupedaddress, noshowpacs]{revtex4}  % for review and submission
\usepackage{graphicx}  % needed for figures
\usepackage{dcolumn}   % needed for some tables
\usepackage{bm}        % for math
\usepackage{amssymb}   % for math
\usepackage{float}
\usepackage{mathtools}
\usepackage{mathrsfs}
\usepackage{amsmath}
\usepackage{bbm}
\usepackage{dsfont}

\newcommand{\fabs}[1]{\left| #1 \right|}

\newcommand{\ket}[1]{\ensuremath{|#1\rangle}}
\newcommand{\bra}[1]{\langle#1|}
\newcommand{\braket}[2]{\langle#1|#2\rangle}

\newcommand{\cPh}{\hat{\begin{cal}P\end{Huge}{cal}}}

\newcommand{\fref}[1]{Fig.~\ref{#1}}

\newcommand{\eref}[1]{Eq.~\eqref{#1}}

\usepackage{ulem} %for \sout
\usepackage{color}

\begin{document}

\title{Quantum Catcher: Trapping and cooling particles using a moving atom diode and\\ an atomic mirror}

\author{Tom Dowdall}
\address{Department of Physics, University College Cork, Cork, Ireland}

\author{Andreas Ruschhaupt}
\address{Department of Physics, University College Cork, Cork, Ireland}

\begin{abstract}
We propose a theoretical scheme for atomic cooling, i.e. the compression of both velocity and position distribution of particles in motion. This is achieved by collisions of the particles with a combination of a moving atomic mirror and a moving atom diode. An atom diode is a unidirectional barrier, i.e. an optical device through which an atom can pass in one direction only. We show that the efficiency of the scheme depends on the trajectory of the diode and the mirror. We examine both the classical and quantum mechanical descriptions of the scheme, along with the numerical simulations to show the efficiency in each case.
\end{abstract}
\maketitle

\section{Introduction}
\label{sec:intro}

One standard cooling technique for neutral atoms is using magneto-optical traps \cite{MOT}. Evaporative cooling of bosons is used for achieving condensates \cite{Ketterle1} and ultracold, spin-polarised Fermi gases are usually cooled to temperatures below the Fermi temperature through sympathetic cooling \cite{Ketterle2}.

Recently another method has been introduced, called single photon cooling \cite{AMDudarev,AndreasMugaRaizen,Raizen}, which allows one to cool atoms and molecules which cannot be handled in a standard way. The method is based on an atom diode or one-way barrier \cite{RuschhauptMuga2,Raizen2}. An atom diode is a device which allows the atom to pass through it only in one direction whereas the atom is reflected if coming from the opposite direction. Such a device has been studied theoretically \cite{AndreasMugaRaizen,RuschhauptMuga3,AndreasMugaRaizen2,AndreasMuga4,AtomDiode} and also experimentally implemented as a realisation of a Maxwell demon \cite{JJThorn,JJThorn2}.

A way of changing or reducing the velocity of particles (which does not necessarily correspond to cooling) is letting particles collide with a moving mirror. An early example is the production of an ultracold beam of neutrons colliding with a moving Ni-surface \cite{Steyerl}. Atomic mirrors can be built using reflection by an evanescent light field
\cite{Adams, Cronin}. Moving such an mirrors for cold atom waves has
been also implemented with a time-modulated, blue-detuned evanescent
light wave propagating along the surface of a glass prism \cite{Steane,Arndt,Szriftgiser}.
More recently, the diffraction of a Bose-Einstein condensate on a vibrating mirror potential
created by a blue-detuned evanescent light field was studied \cite{Colombe}
and the reflection of an atomic cloud from an optical barrier of a blue-detuned beam
was used to study first-order and second-order catastrophes in the cloud density \cite{Rosenblum}.
Even Rb atoms which fall on a magnetic mirror have been examined \cite{Roach} and
Rb atoms have even been stopped using a moving magnetic mirror \cite{Reinaudi}.
Furthermore solid atomic mirrors have been used for focusing neutral atomic and molecular beams
\cite{Holst, Fladischer, Anemone}.
Si-crystals on a spinning rotor have been used as a solid atomic mirrors to slow down
beams of Helium atoms \cite{Libson, Narevicius}.

A stream of particles can be slowed by collision with a moving mirror travelling in the same direction as the particles.
One limitation of standard settings at present is  
that for a fixed mirror velocity only pulses of particles with a specific and well defined initial velocity are stopped.   
In \cite{SSchmidt}, it was shown that by designing a particular trajectory for the mirror it is even possible
to stop a pulse in which the initial velocities are broadly distributed or possibly unknown.
But slowing an ensemble of atoms solely with one mirror of course does not result in phase-space compression. In order to achieve this, we introduce a required irreversible step.

In this work  we develop a scheme to cool (\textit{i.e} compress in phase space) a travelling cloud of particles.
This is done by combining the idea of a moving mirror with an irreversible atom diode also in motion.

In the next section, we present and investigate our cooling method, first in an idealised classical setting, i.e. assuming
a point-particle with classical motion. In Section \ref{sec:quantum}, we discuss a quantum-mechanical implementation
of our cooling scheme. The paper ends with a conclusion.

%---------------------------------------------------------------------------------
%------------------------ Classical Catcher ----------------------------------------
%---------------------------------------------------------------------------------

\section{Cooling classical particles with diode and mirror}
\label{sec:classical}

First we shall investigate a classical scheme for achieving our goals before moving on to a full quantum treatment of the problem.
We assume classical point particles and restrict the scenario to a one-dimension motion.
The setting consists of two main objects: a moving atomic mirror potential and an atom diode.
The particles move freely between the collisions with these two objects.
Let us start by reviewing properties of a single moving atomic mirror potential.

\subsection{Elastic collision stopping a single particle with moving mirror}

A collision between a number bodies is called elastic if there is no loss of mechanical energy during the collision. With this in mind consider the collision of a particle (moving with velocity $v_{p}$) with a moving mirror (with velocity $v_{m}$). The velocity of the particle after the elastic collision is given by
\begin{equation}
v_{f} = 2 v_{m} - v_{p}.
\label{eq:M-D_Col}
\end{equation}
It is immediately apparent that if we let $v_{m} = \frac{v_{p}}{2}$ the particle is stopped instantly by the collision. We can see that in particular, if a particle has trajectory $x(t) = v_{p} t$, the trajectory of the mirror is $x_{m}(t)$ and the collision occurs at time $t_{c}$ then we have
\begin{eqnarray}
\dfrac{d x_{m}}{dt}\bigg|_{t_{c}} = \dfrac{v_{p}}{2} = \dfrac{x_{m}(t_{c})}{2 t_{c}}.
\end{eqnarray}
We require that the same mirror trajectory should stop all particles independent of their velocity $v_p > 0$, i.e. the previous equation should be fulfilled for all $t_c > 0$.
This ordinary differential equation (with $t_{c}$ replaced by $t$) has then a solution $x_{m}(t) = \alpha\sqrt{t}$ with $\alpha > 0$. This trap trajectory has been  explored in \cite{SSchmidt}, where it stops particles of arbitrary velocity. Unfortunately, these particles can be completely delocalised in space and thus no real cooling (i.e. phase space compression) is achieved with just a single atomic mirror.

%----------------- Figure 1 -------------------------------------------
\begin{figure}[t]
\begin{center}
\includegraphics[width = 0.95 \linewidth]{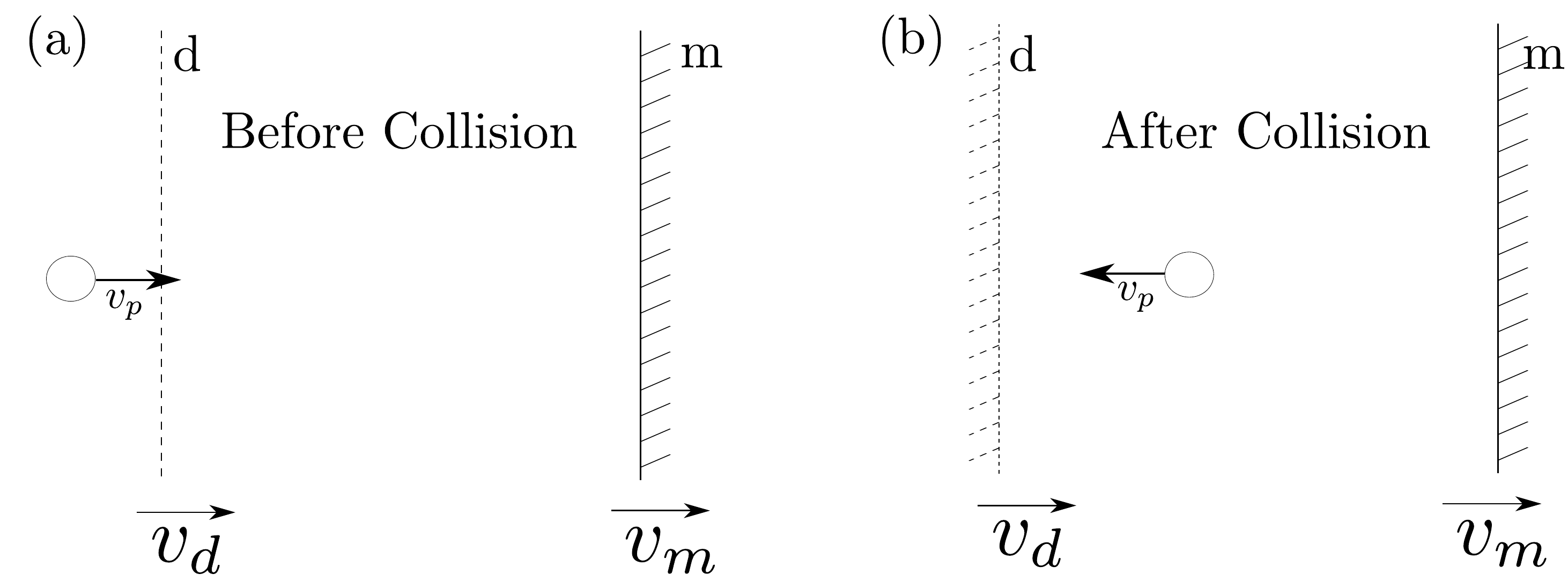}\\
(c) \includegraphics[width = 0.75 \linewidth]{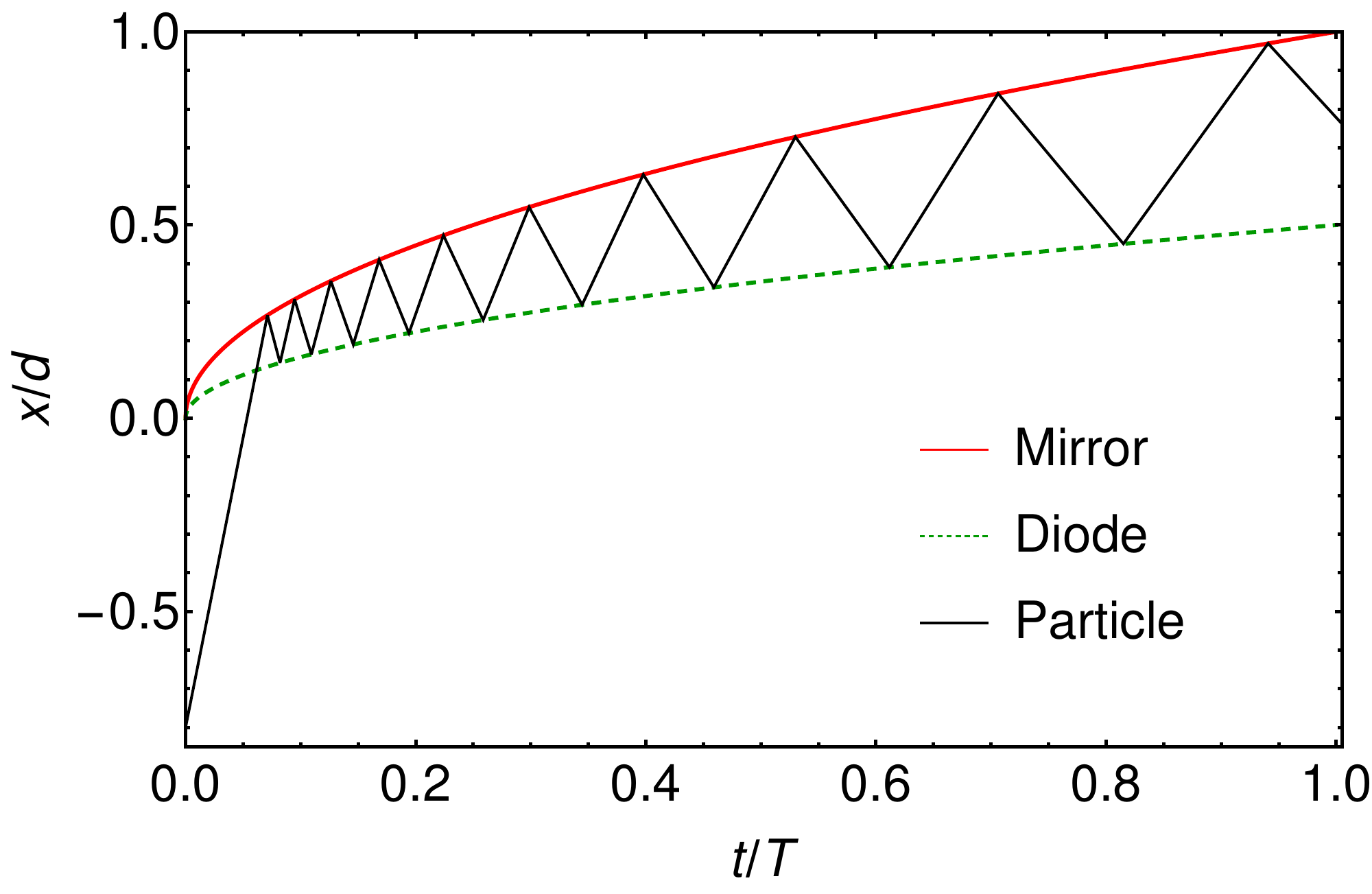}\\
(d) \includegraphics[width = 0.75 \linewidth]{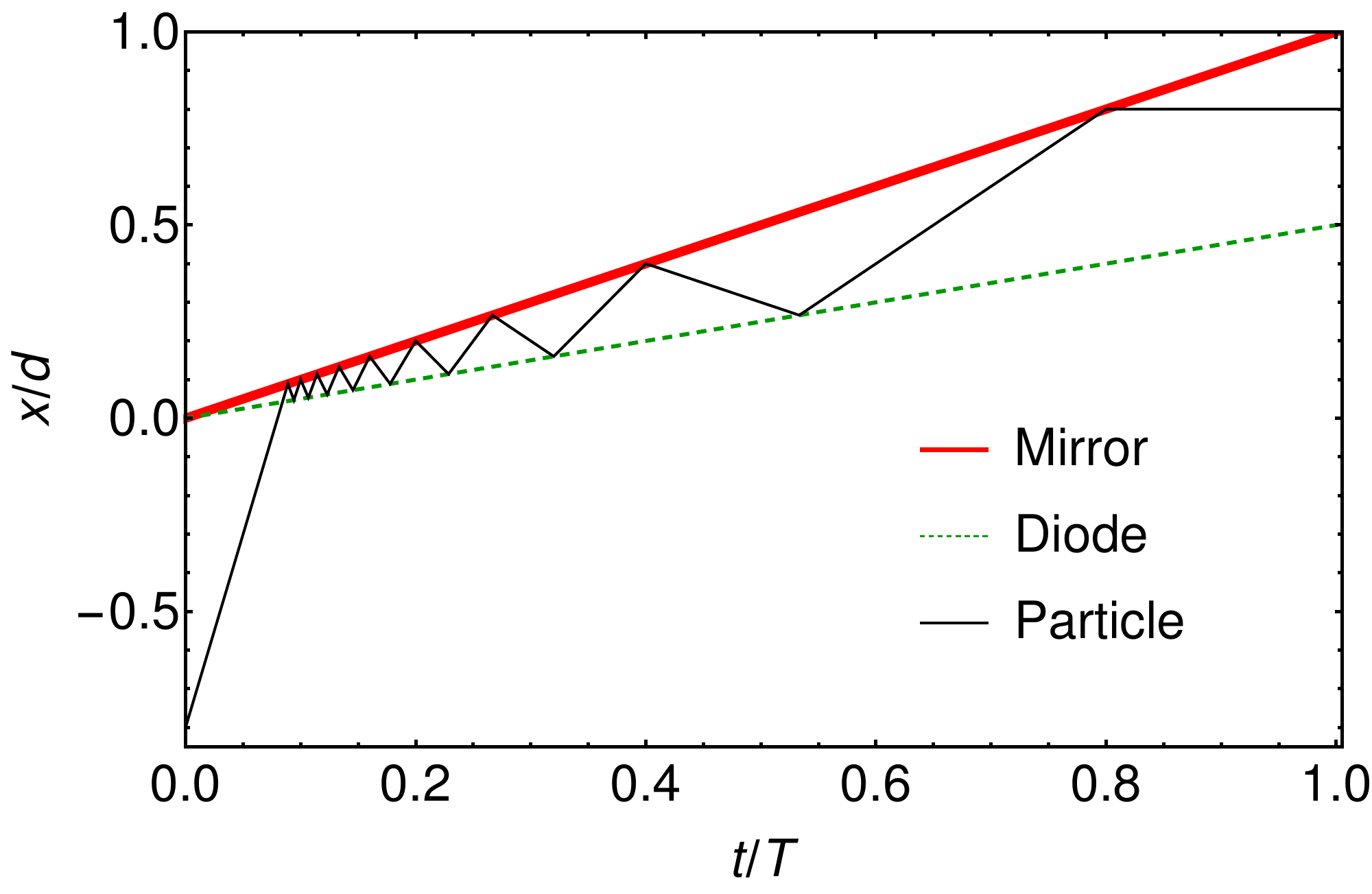}\\
\caption{
(a) and (b) Diode-mirror setting: A particle approaches the moving diode-mirror system; it can enter one way through the diode in (a) but in (b) from the other direction the diode behaves as another mirror travelling at a different velocity.
(c) and (d) Motion of the atom diode and the mirror with trajectories (c) $\sim \sqrt{t}$ and (d) $\sim t$.
}
\label{fig:md}
\end{center}
\end{figure}
%---------------------------------------------------------------------

\subsection{Cooling with atom diode and atomic mirror}

In this paper we propose a slightly different approach. Instead of attempting to stop the particles we demonstrate a method for cooling them.

A schematic of our setting is seen in \fref{fig:md} (a) and (b): it consists of an atom diode (d) shown here on the left and a mirror potential (m) on the right, moving with velocities $v_d$ and $v_m$.
Let us consider a single particle incident on the diode from one direction (here from the left to the right) which passes through (\fref{fig:md}a). The particle is then reflected by the mirror as a result its absolute velocity is reduced.
However in the next collision the particle is reflected by the diode which now acts as an atomic mirror (\fref{fig:md}b).

In \fref{fig:md} (c) and (d) this idea demonstrated again; the particle incident from the below can pass through the barrier but this particle, when it is then travelling downwards, is reflected by the diode.
This traps the particle in between the two objects. According to \eref{eq:M-D_Col} every time the particle collides with the mirror it experiences a reduction in velocity and every time
the particle is reflected by the diode, its velocity is increased. Since the mirrors is travelling at a faster velocity than the diode, there is an overall reduction in velocity after
two collisions. The absolute velocities the particle continue to slow down until the particle is not travelling fast enough to collide with the mirror.
Because the setting confines the particle and the collisions between the particle and the moving diode/mirror slow down the particle, through continued collisions inside the diode-mirror trap a cooling can be achieved.

This idea was first proposed in \cite{Schmidt_PhD} where both diode and mirror travel with the same velocity $\sim 1/\sqrt{t}$ but they are displaced by a constant distance. With these trajectories a slight compression in velocity has been achieved.

In this work, we show that the efficiency depends strongly on the trajectories of diode and mirror. By considering different trajectories, we show that significant phase space compression can be achieved. Motivated by the Section II A, we first consider a square-root scheme where the trajectories of diode (d) and mirror (m) are
\begin{eqnarray}
x_{d}(t)= \alpha_{d} \sqrt{t}, \hspace{1 cm}
x_{m}(t)= \alpha_{m} \sqrt{t},
\end{eqnarray}
with $\alpha_m > \alpha_d$, see also \fref{fig:md} (c).

Alternatively, we consider a linear scheme where the trajectories of diode (d) and mirror (m) are
\begin{eqnarray}
x_{d}(t)= v_{d} t, \hspace{1 cm}
x_{m}(t)= v_{m} t, 
\end{eqnarray}
with $v_m > v_d$, see also \fref{fig:md} (d).
As it will turn out later that the linear scheme is more advantageous than the square root scheme,
we derive some general formulas and properties for the linear scheme first.

\subsection{Properties of the linear scheme}
\begin{figure}[t]
\begin{center}
\includegraphics[width = 0.8 \linewidth]{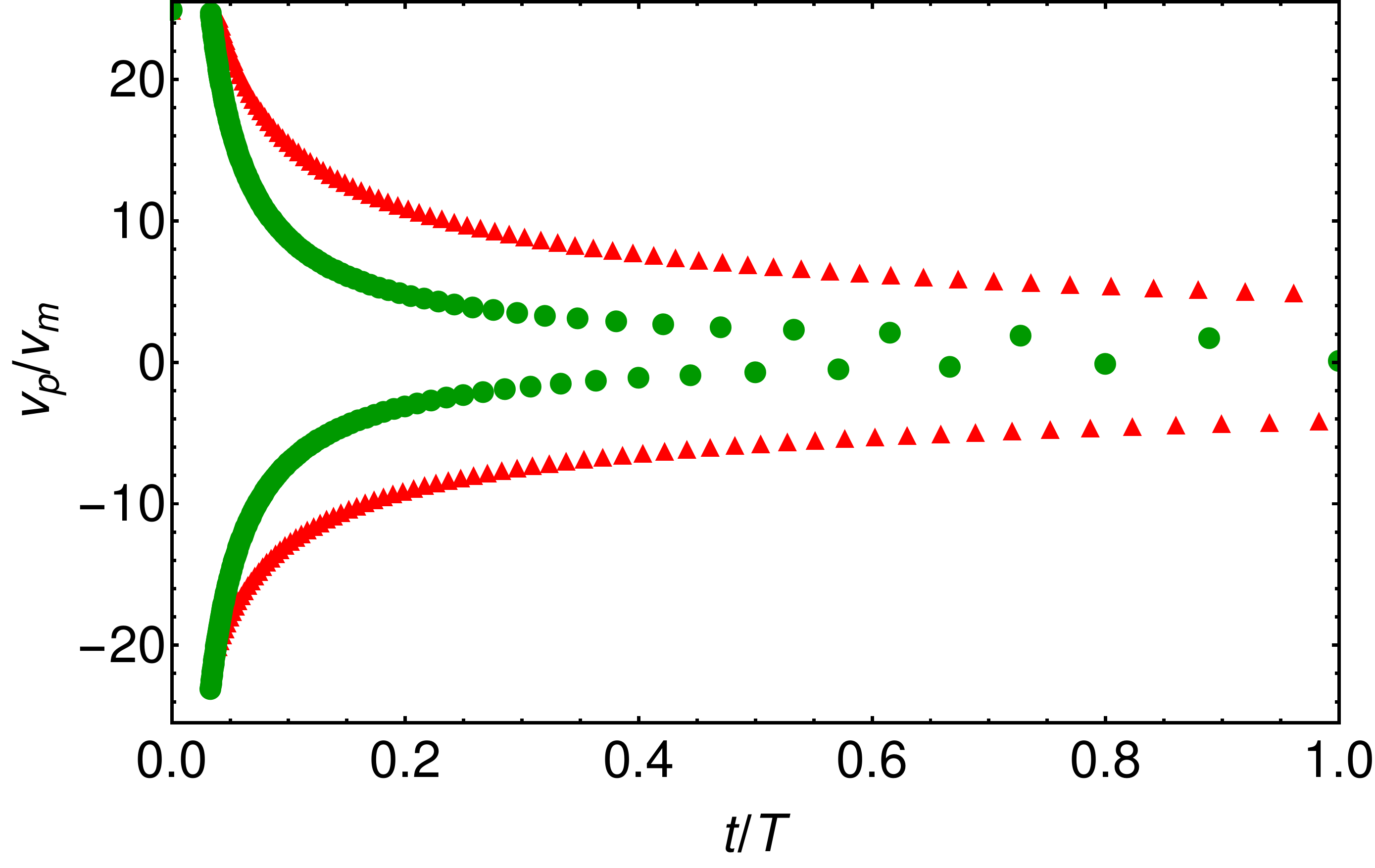}
\caption{Classical setting: graph of the velocity of the particle as a function of time, each symbol indicates the velocity of the particle after a collision; parameters for linear scheme (green dots): $v_{m} = d/T$ , $v_{d} = 0.9 v_m$;
parameters for square root scheme (red triangles): $\alpha_{m} = v_m \sqrt{T}$, $\alpha_{d} = v_d \sqrt{T}$.\label{fig:v(t)}}
\end{center}
\end{figure}

In the linear case, there is an explicit formula for the velocity of the classical particle after the $n^{th}$ collisions, namely
\begin{equation}
v_{n} = \left\lbrace 
\begin{array}{cc}
n  (v_{d} - v_{m}) + v_{i}\hspace{5 mm} & n \text{ even}\\
(n-1) (v_{m} - v_{d}) + 2 v_{m} - v_{i} \hspace{5 mm}& n \text{ odd} 
\end{array}\right.
\label{eq:v(n)}
\end{equation}
where even $n$ corresponds to the velocity after a diode collision and odd $n$ corresponds to the velocity after a mirror collision.
We can also write down an expression for the corresponding time $t_{n}$ for which the $n^{th}$ collision happens 
\begin{equation}
t_{n} = \dfrac{x_{i}}{v_{m} - v_{i}} \left(\prod_{k \text{ even}}^{n-1} \dfrac{v_{k} - v_{d}}{v_{k} - v_{m}}\right) \left(\prod_{l\text{ odd}}^{n-1} \dfrac{v_{l} - v_{m}}{v_{l} - v_{d}}\right).
\label{eq:t(n)}
\end{equation}
We can use \eref{eq:v(n)} to calculate the maximum number of collisions (if there is no further time restriction):
After the last collision ($n=n_{max}$), we have $v_{d} \leq v_{n_{max}} \leq v_{m}$. From this, it follows:
\begin{eqnarray}
v_{d} \leq &v_{i} - n_{max} \Delta v_{md}& \leq v_{m},\nonumber\\
\dfrac{v_{i} - v_{m}}{\Delta v_{md}} \leq &n_{max}& \leq \dfrac{v_{i} - v_{d}}{\Delta v_{md}}, \nonumber\\
r - 1 \leq &n_{max}& \leq r
\end{eqnarray}
where $\Delta v_{md} = v_m - v_d > 0$ and $r = \dfrac{v_{i} - v_{d}}{\Delta v_{md}}$.
For an even $n$, with $1 < n \le n_{max}$, it follows therefore that $n \le r$ and therefore
\begin{eqnarray}
v_{i} - n \Delta v_{md} &\geq& (n-2) \Delta v_{md} + 2 v_{m} - v_{i},\nonumber\\
v_{n} &\geq& v_{n-1}.
\end{eqnarray}
From \eref{eq:v(n)}, it also follows immediately that
\begin{eqnarray}
v_{n} - v_{n-2} = \left\lbrace 
\begin{array}{cl}
-2 \Delta v_{md} < 0\hspace{1 cm} & n\text{ even}\\
2 \Delta v_{md} > 0\hspace{1 cm} & n\text{ odd}
\end{array}\right. .
\label{eq:brach}
\end{eqnarray}
From these last results, it is seen that the velocities after diode collisions ($n$ even) are decreasing with increasing number of collisions
and the velocities after mirror collisions ($n$ odd) are increasing with increasing number of collisions.
It also follows that all velocities after diode collisions ($v_n$ with $n$ even) are always larger than the velocities after mirror collisions
($v_n$ with $n$ odd). We will see all these general properties also in the example in \fref{fig:v(t)} discussed in the next section.

% ---------------- Figure 2 ------------------------------------------------
% --------------------------------------------------------------------------

%----------------- Figure 3 ------------------------------------------------
\begin{figure}[t]
\begin{center}
(a) \includegraphics[width = 0.75\linewidth]{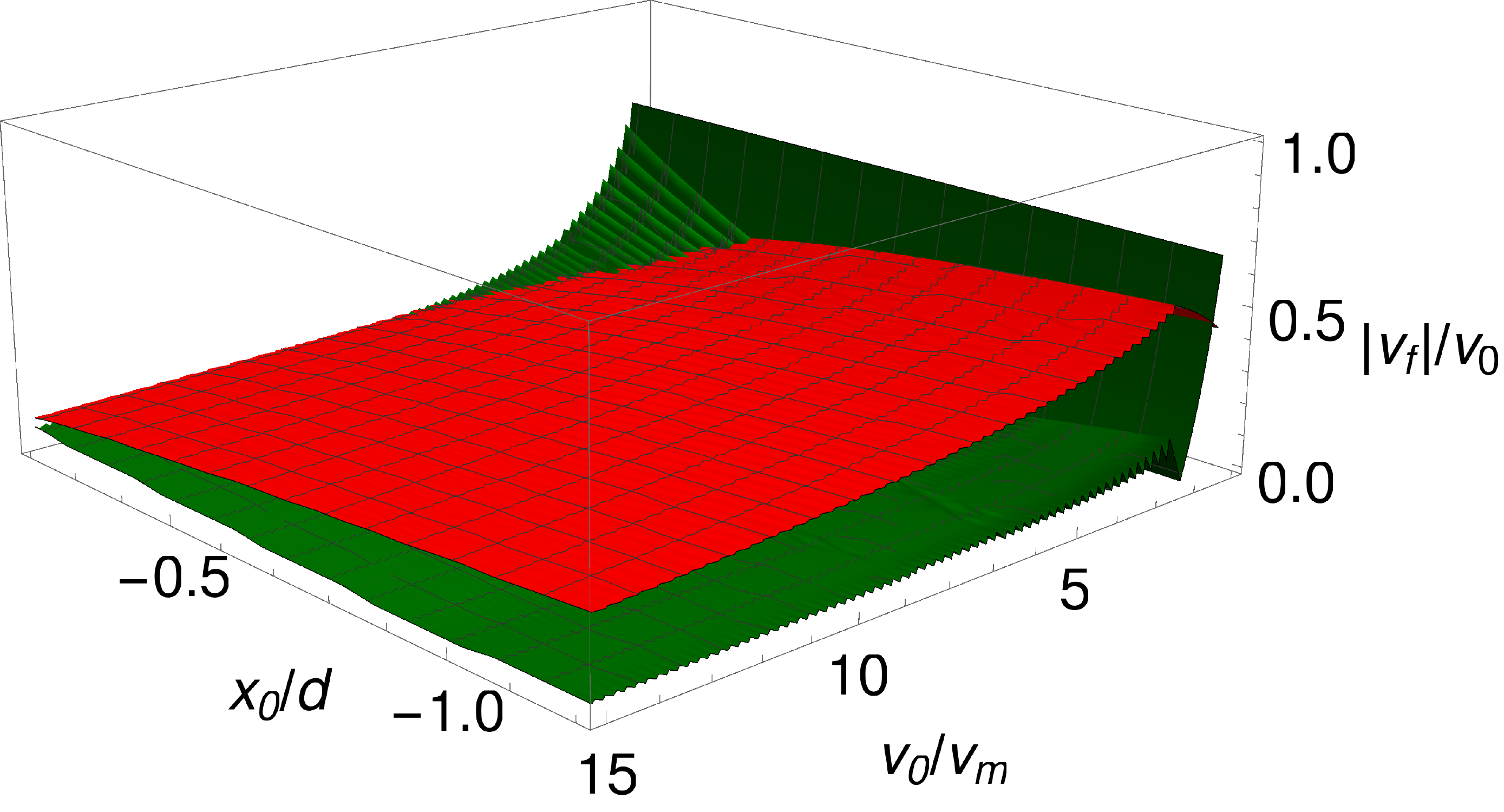}

(b) \includegraphics[width = 0.75\linewidth]{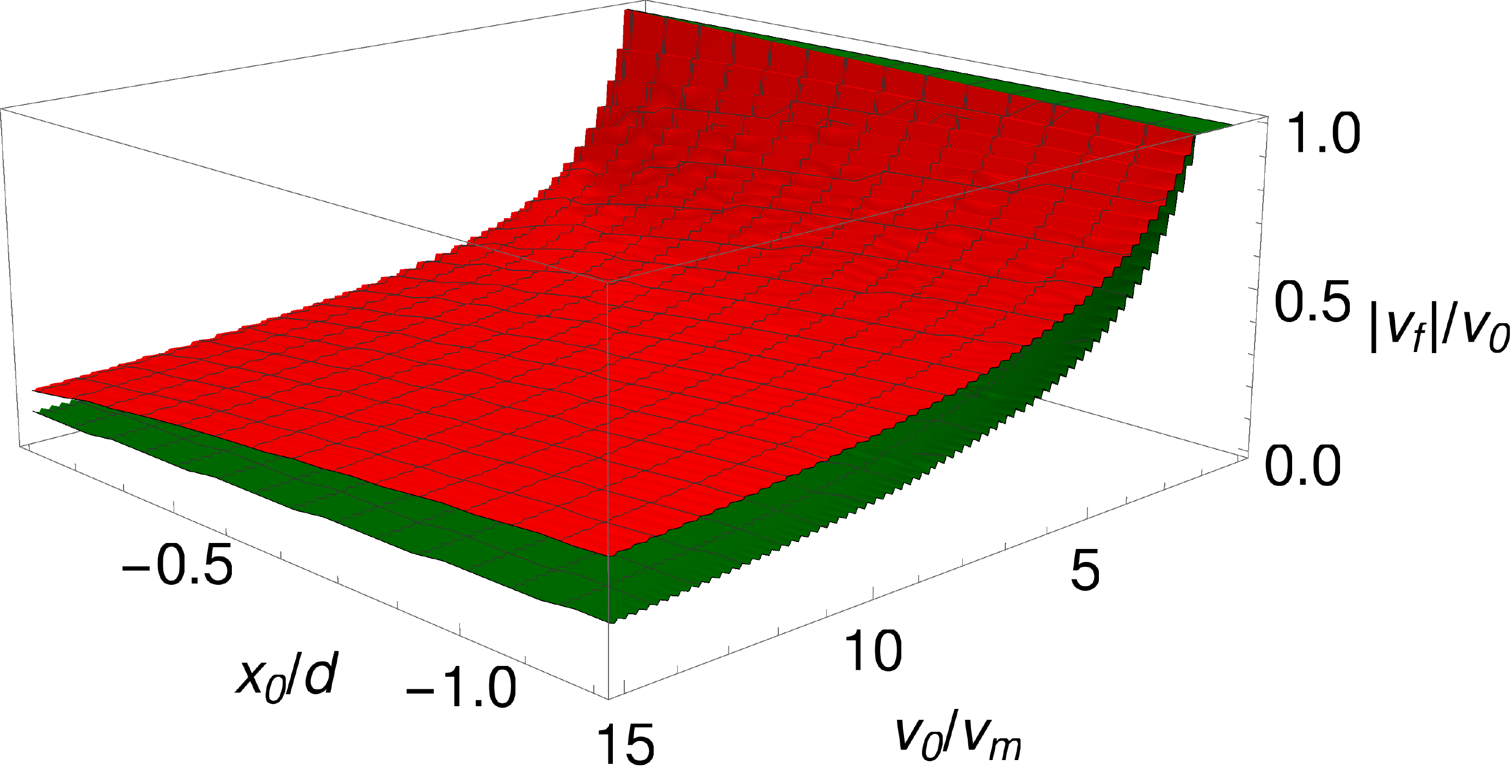}
\caption{Classical setting: plot of $\frac{\fabs{v_f}}{v_{0}}$ versus initial particle velocity $v_{0}$ and initial particle position $x_{0}$:
(a) velocity $v_{f}$ after the last collision with the mirror, (b) velocity $v_{f}$ after the last collision with the diode.
Linear scheme (green, lower planes) and square root scheme (red, higher planes), other parameters are the same as in \fref{fig:v(t)}.\label{classicalcomp} 
}
\end{center}
\end{figure}
%----------------------------------------------------------------------------

\subsection{Comparison of the square-root and linear schemes for a single particle}
Let $v_m = d/T$ where $d$ is the final position of the mirror and $T$ is the total time, $v_{m}$ is also the velocity  of the mirror in the linear scheme.
For comparison, we chose $\alpha_{d/m}=v_{d/m} \sqrt{T}$ in the square-root scheme in such a way that the initial and final position of diode and mirror is the same in both schemes.

In \fref{fig:v(t)}, the velocity of the particle $v_p$ after a collision is shown versus time, for the square-root scheme as well as for the linear scheme.
We see the velocity of the particle in the trap tends towards $v_{d} \leq v_{p}(t) \leq v_{m}$ for larger $t$; furthermore the particle is localised $x_{d}(t) \leq x_{p}(t) \leq x_{m(t)}$.
We see this behaviour in the linear case and in the case of the square root;
however we do not see the same level of velocity reduction in \fref{fig:v(t)} in the square-root case as in the linear case:
the reducing of the velocity occurs in the linear trap on a much shorter timescale than that of the square root trap (it takes much longer to achieve the same reduction in velocity for the square-root trap).

If we consider again the linear case in \fref{fig:v(t)}, then we will also see all the general properties derived in Section II: the upper branch (corresponds to $n$ even, i.e.
velocities after diode collisions) is decreasing with increasing time (which correspond to increasing number of collisions), the lower branch (corresponds to $n$ odd, i.e.
velocities after mirror collisions) is increasing with increasing time (which correspond to increasing number of collisions) and the upper branch is always above the lower branch.

The ratio between final particle velocity after the last mirror resp. diode collision and the initial particle velocity is shown in \fref{classicalcomp}. We see from $|v_{f}|/v_{0} < 1$ that we have achieved a reduction in velocity.
We can compare this relative performance of the square root and linear schemes.
We see the linear scheme is much more successful for reducing final velocity ($\frac{\fabs{v_{f}}}{v_{0}}$ displayed) than the square root scheme.
The surfaces begin to approach each other when the particle that is travelling slowly and starts close to the diode-mirror system.
This is because a slow travelling particle is less likely to collide with the the diode-mirror system and so is less likely to have achieved any velocity reduction.

From \fref{classicalcomp} we expect that by instead sending in not a single particle with well-defined position and velocity but a particle or an ensemble with a probability
distribution of velocity and position, we achieve the cooling desired. This is examined in the following.

% ---------------- Figure 4 --------------------------------------------------------
\begin{figure}[t]
\begin{center}
\includegraphics[width=0.75\linewidth]{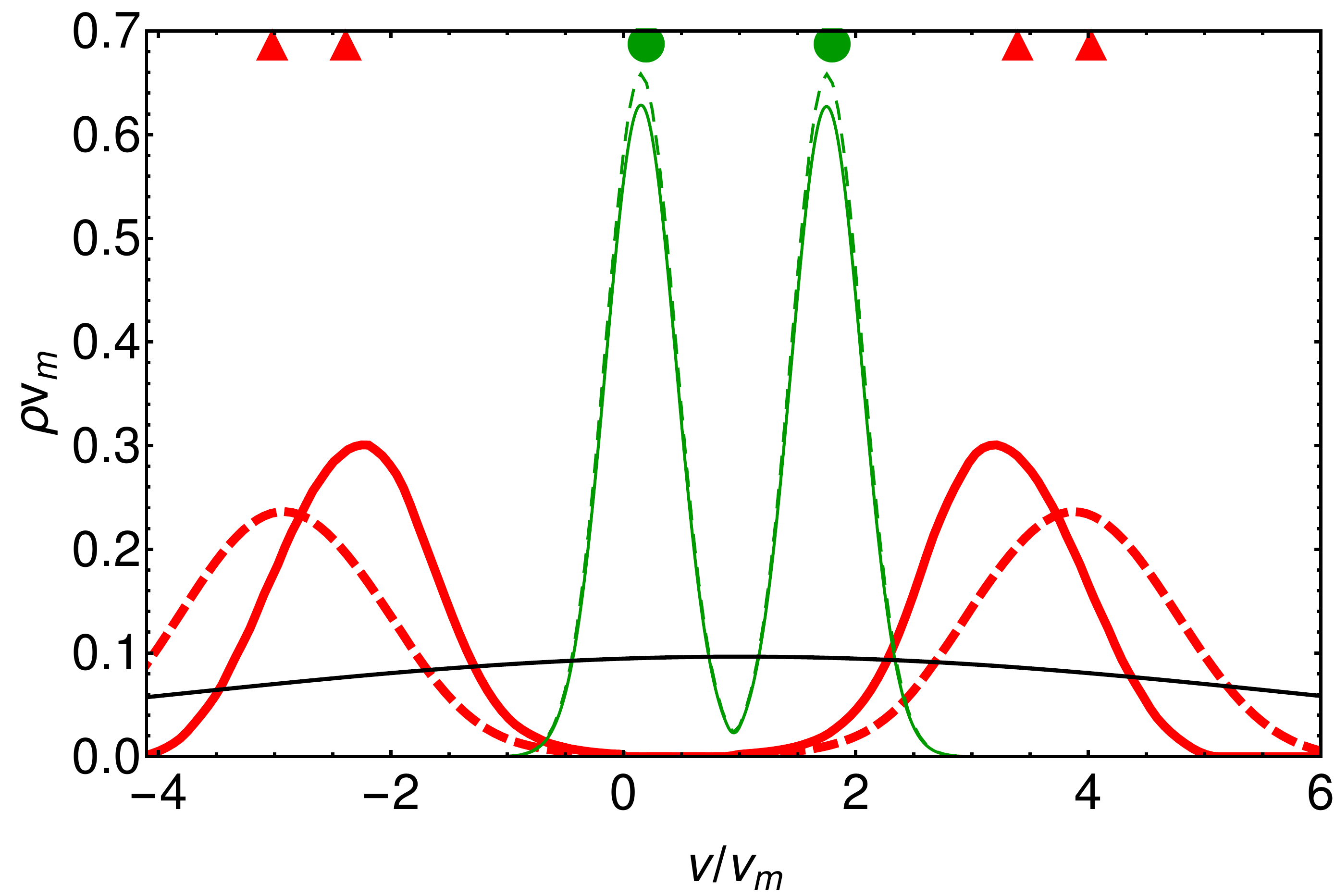}
\caption{Classical setting: comparison between velocity distribution using the linear and square root schemes: initial velocity distribution for both schemes (shifted, black, lowest broad distribution),
final velocity distribution: for the square-root scheme: $v_0 = 10 v_m$ (red, thick, solid line), $v_0 = 15 v_m$ (red, thick, dashed line);
for the linear scheme: $v_0 = 10 v_m$ (green, thin, solid line), $v_0 = 15 v_m$ (green, thin, dashed line). The dots above the plots correspond to a single particle simulation with initial velocity $v_{0}$ and initial position $x_{0}$; other parameters: $x_0 = -0.8 d, \Delta x = 0.1 d, \Delta v = 5 v_m$;
other parameters are the same as in \fref{fig:v(t)}. \label{sqrt_class_comp}
}
\end{center}
\end{figure}
% ---------------------------------------------------------------------------------

% ---------------- Figure 5----------------------------------------------
\begin{figure}[t]
\begin{center}
(a)\includegraphics[width=0.7\linewidth]{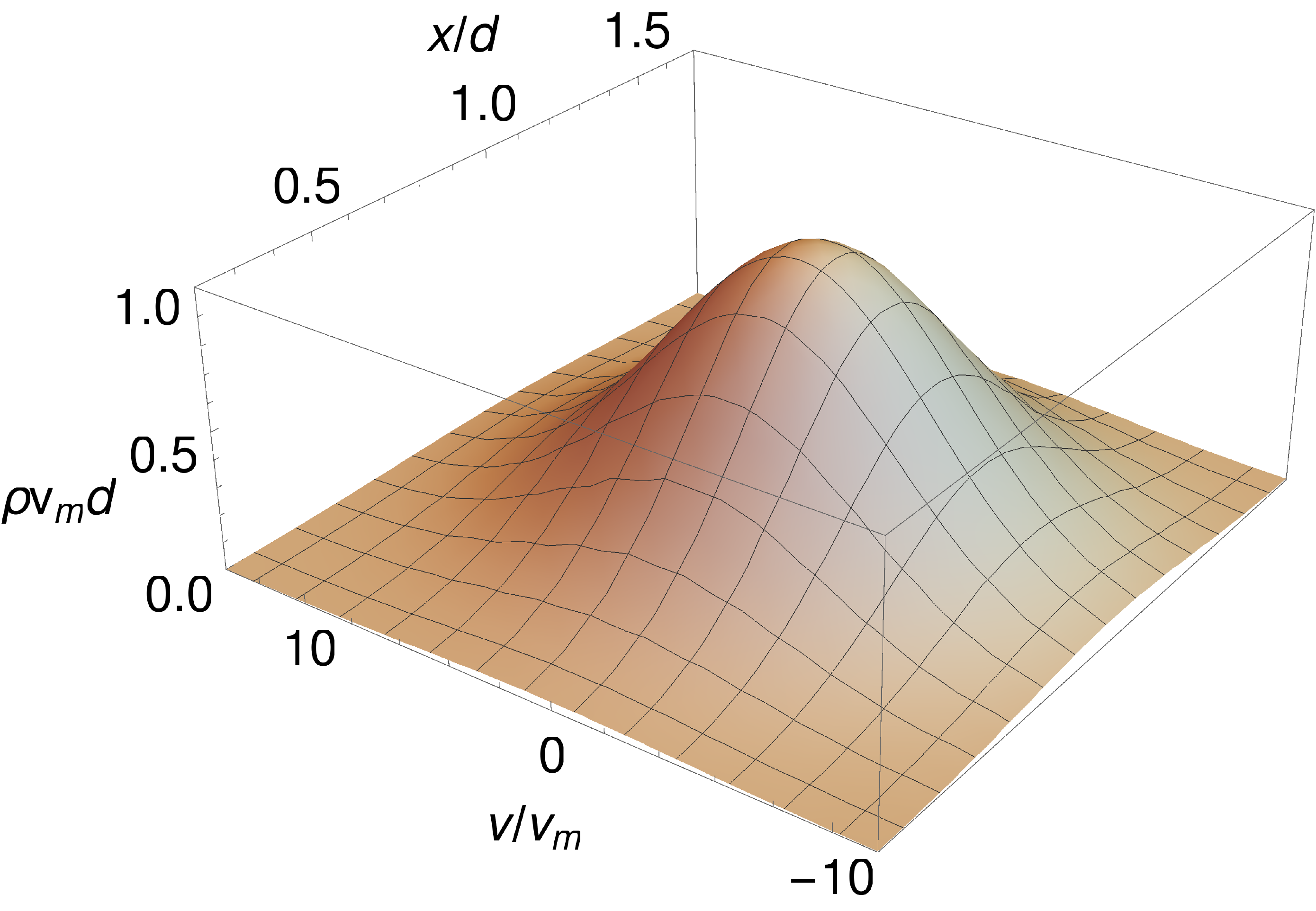}\\[1cm]
(b)\includegraphics[width=0.7\linewidth]{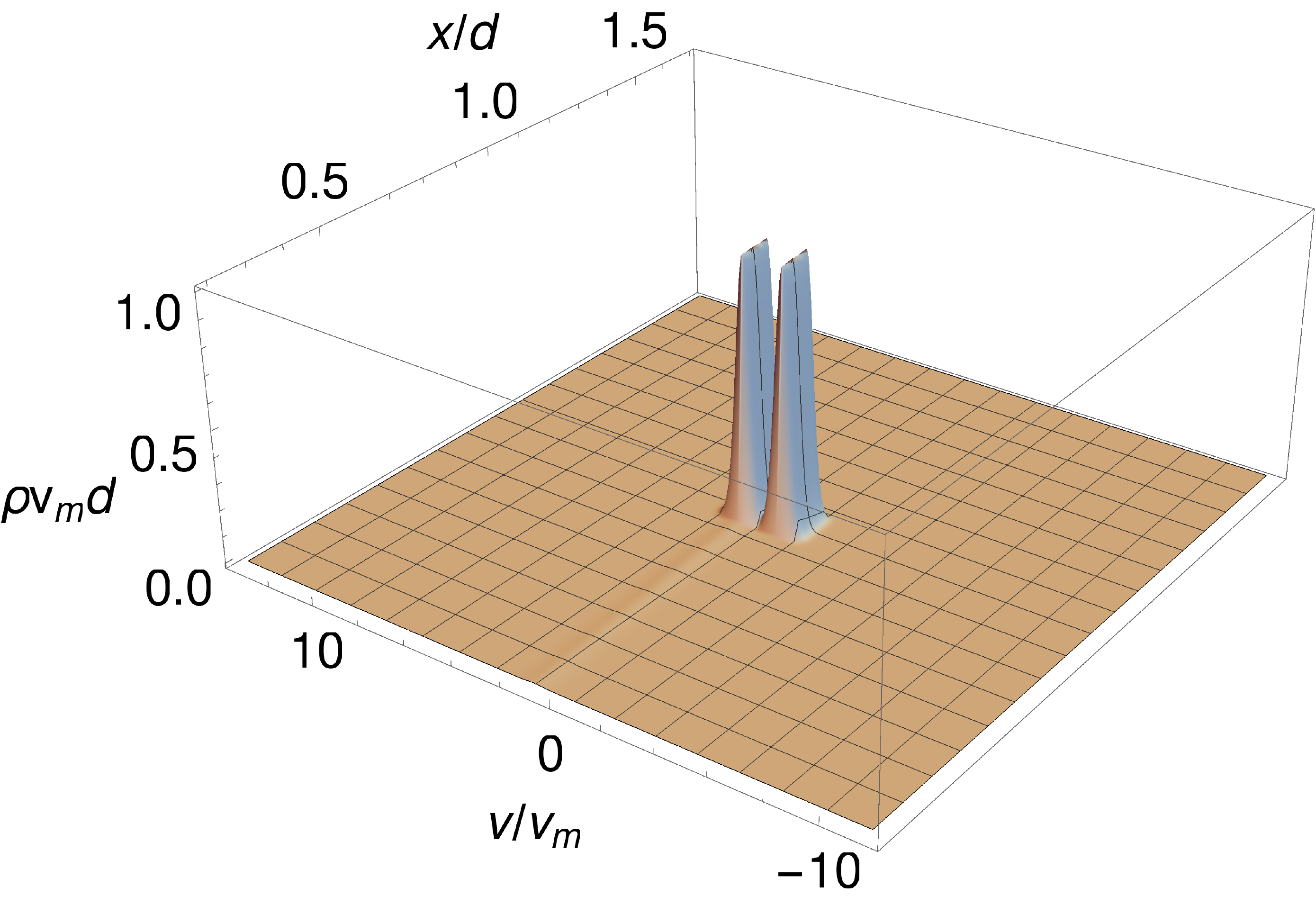} 
\caption{Classical setting: (a) shifted initial distribution $\rho(0,x,v)$ and (b) final distribution $\rho(T,x,v)$. Both distributions are scaled such that their maximum is one. Linear scheme,
$v_0 = 10 v_{m}$, other parameters as in \fref{sqrt_class_comp}.
\label{3dplot}}
\end{center}
\end{figure}
% ------------------------------------------------------------------

\subsection{Compression in classical phase space}
We now discuss the more general case where we have a cloud of non-interacting particles characterised by some probability density $\rho (t,x,v)$. 
In particular we look at a Gaussian initial distribution given by
\begin{eqnarray}
\rho(0, x, v) = \dfrac{1}{2 \pi \Delta v \Delta x} e^{-\left[\left(\frac{x - x_{0}}{2 \Delta x }\right)^{2} + \left(\frac{v - v_{0}}{2 \Delta v}\right)^{2}\right]}.
\end{eqnarray}
We calculate the final probability distribution at time $t=T$, $\rho (T,x,v)$ for the linear and square root schemes and compare the ability in each case to cool the cloud.
In \fref{sqrt_class_comp} this comparison between the initial and final velocity distributions ($\rho(t,v) =  \int dx \rho(t,x,v)$ for $t = 0,T$) is shown and
we see that both schemes achieve a reduction in velocity. The linear scheme however achieves a greater reduction in velocity than the square root one similar to the single particle case shown in \fref{fig:v(t)} and \fref{classicalcomp}.
It is interesting that the final velocity distribution is independent of the initial average velocity $v_0$ for the linear scheme.
The dots in \fref{sqrt_class_comp} correspond to the final velocities after the mirror collision resp. diode collision which are achieved if we consider a single particle in the diode-mirror system with $v_{0}$ and $x_{0}$ being the average velocity and position of the ensemble.
We find that the positions of the peaks correspond approximately to these velocities.
To underline the compression in phase space, the initial and final distribution $\rho(0,x,v)$ resp. $\rho(T,x,v)$ is shown in \fref{3dplot} for the linear scheme.
For clarification, both distributions are shown scaled such that their maximum is one and the initial distribution is also shifted.
It can be clearly seen that the cooling resp. compression in phase space is achieved.

We have shown that the efficiency depends strongly on the trajectories of atom diode and atomic mirror. It turns out that the linear scheme
is much more efficient that the square-root scheme in the classical setting. Therefore, we will consider now solely the linear scheme in a quantum setting.

%---------------------------------------------------------------------------------
%------------------------ Quantum Catcher ----------------------------------------
%---------------------------------------------------------------------------------

\section{Quantum Catcher}
\label{sec:quantum}
Inspired by the preliminary and promising classical results, we would like to consider if such a similar cooling is possible using a quantum mechanical treatment.
We again consider a single quantum particle moving in one dimension. 
We want the quantum diode-mirror system to operate similarly to the classical case;
we expect however differences as there will be quantum effects and the dependence on mass in the Schr\"odinger equation.

\subsection{Implementing a quantum atom diode and mirror}
\begin{figure}[t]
\begin{center}
\includegraphics[width=0.8\linewidth]{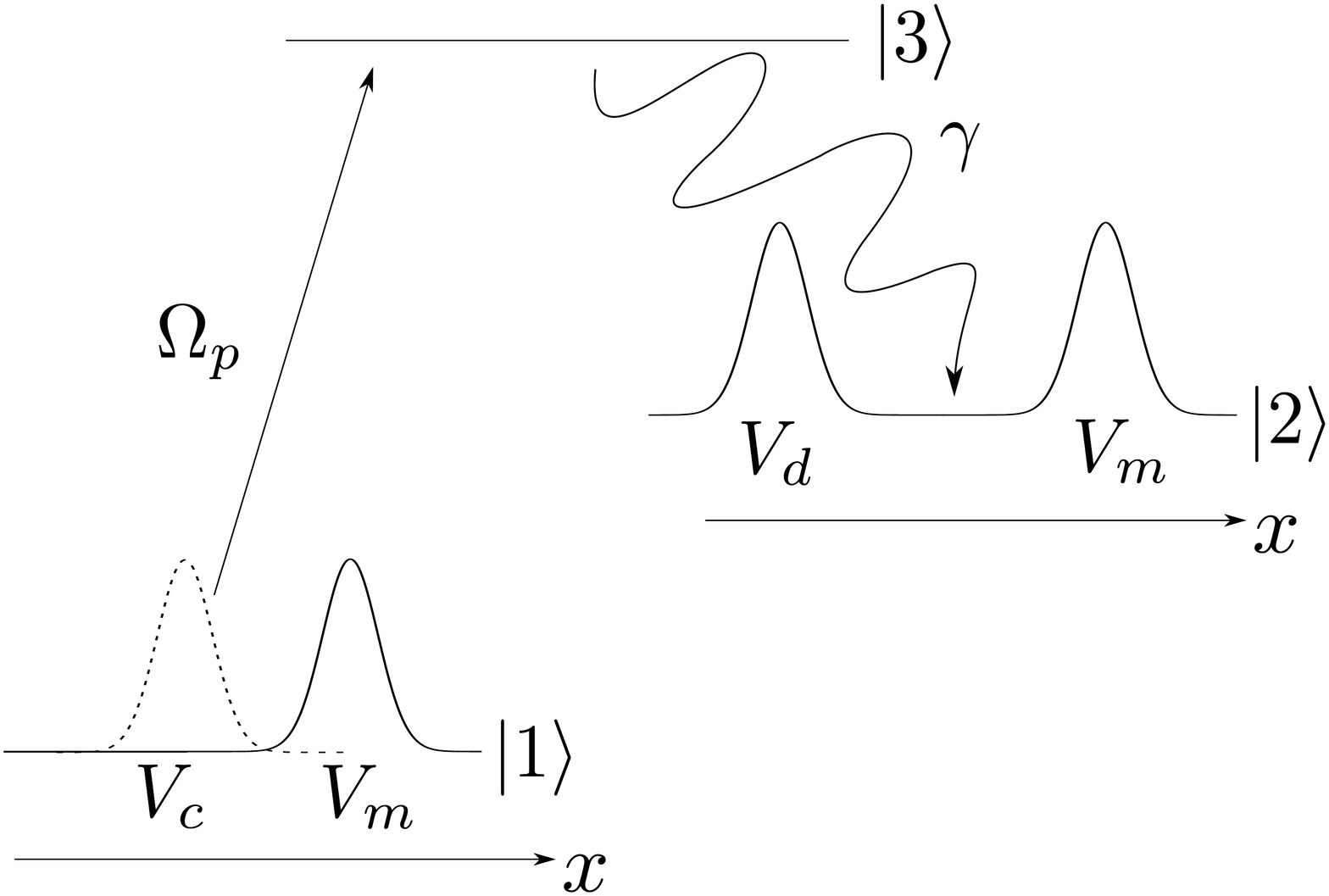}
\caption{Quantum atom diode and atomic mirror scheme.}
\label{fig:atom-diode}
\end{center}
\end{figure}

While the reflection mirror can be realised for example in experiments by an optical potential, the implementation
of an atom diode is less straightforward.
A theoretical proposal for such a diode is found for example in \cite{AtomDiode} and a similar one (see \fref{fig:atom-diode}) we use throughout the remaining paper.
We start with the mirror potential $V_{m}(x)$ which acts on the atom independent of whether it's in state $\ket{1}$ or $\ket{2}$. For implementation of the atom diode, we assume a coupling between levels $\ket{1}$ and $\ket{3}$ with a Rabi frequency $\Omega_{p}(x)$. State $\ket{3}$ decays quickly with decay constant $\gamma$ to the stable state $\ket{2}$. Finally there is a state selective potential $V_{d}(x)$ (placed on the left hand side of $V_{m}(x)$) which effects the atom only if it is in state $\ket{2}$. Assume the particle is now incident from the left in state $\ket{1}$ it is then pumped to state $\ket{3}$ where it decays to state $\ket{2}$. There it is trapped between the two potentials $V_{d}(x)$ and $V_{m}(x)$.
% ---------------- Figure 6 ----------------------------------------
% ------------------------------------------------------------------

The master equation for the three level diode-mirror system described above (neglecting recoil) is
\begin{eqnarray}
\dfrac{\partial}{\partial t} \rho (t) =&  - \frac{i}{\hbar} \left[\hat{H}_{3L},\rho(t) \right]_{-} - \dfrac{\gamma}{2} \left\{ \rho(t) \ket{3}\bra{3} + \ket{3}\bra{3} \rho(t) \right\} \nonumber\\
&+ \gamma \ket{2} \bra{3}\rho(t)\ket{3}\bra{2}.
\label{eq:mastereq}
\end{eqnarray}
The Hamiltonian is
\begin{eqnarray}
\lefteqn{\hat{H}_{3L}  =  - \frac{\hbar^2}{2m} \frac{\partial^2}{\partial x^2}} && \\
&& + \left(\begin{array}{ccc}
V_{m}(x,t) & 0& \hbar \Omega_{p}(x,t)/2 \\
0 & V_{d}(x,t) + V_{m}(x,t) & 0 \\
\hbar \Omega_{p}(x,t)/2 & 0 & 0\end{array}\right). \nonumber
\end{eqnarray}
The situation is quite different from the classical case because here the probability density depends on the mass $m$ of the particle chosen.

At initial time $t=0$, we start in a pure state and the initial wavefunction of the particle is a Gaussian (not necessarily a minimum-uncertainty product one)
\begin{eqnarray}
\lefteqn{\psi_0(x) = A \times} && \\
&&\exp\bigg\{ - \frac{1}{1+i c} \bigg(\frac{m^2 \Delta v^2}{\hbar^2} \left(x-x_{0}\right)^2
 + i \frac{m v_0}{\hbar} (x-x_{0})\bigg)\bigg\} \nonumber
\end{eqnarray}
where $c = \sqrt{\frac{\Delta x^2 m^2 \Delta v^2}{\hbar^2} - \frac{1}{4}}$ and $A$ is a normalisation constant.
Note that $c \ge 0$ due to the Heisenberg uncertainty relation.

We use the quantum jump/trajectory approach \cite{QuantumJump1,QuantumJump2,QuantumTrajectory1,QuantumTrajectory2}
to solve the above 1D master equation \eqref{eq:mastereq} numerically. 
In the quantum-jump approach, the master equation \eqref{eq:mastereq}  is solved by averaging over ``trajectories'' with time intervals
in which the wave function evolves with the conditional Hamiltonian interrupted by random jumps (decay events).
In the dynamics before the first spontaneous photon emission, we assume that the quenching laser $\Omega_p$ and the decay
can be approximated by an effective complex potential $-i V_c(x - x_c(t)) = -i\frac{\hbar\Omega_p(x - x_c(t))^2}{2\gamma}$. To be more explicit, before
the jump we model our effective Hamiltonian by
\begin{eqnarray}
\hat{H}_{A} = - \frac{\hbar^2}{2m} \frac{\partial^2}{\partial x^2} + V_{m}(x - x_m (t)) - i  V_{c}(x - x_c(t))
\end{eqnarray}
and after the jump we model our Hamiltonian by
\begin{eqnarray}
\hat{H}_{B} = - \frac{\hbar^2}{2m} \frac{\partial^2}{\partial x^2} + V_{m}(x - x_m(t)) + V_{d}(x - x_d(t))
\end{eqnarray}
where 
\begin{equation}
V_{d/m}(x) = V_{0,d/m} e^{\frac{-x^{2}}{2\sigma_{d/m}}},\;
V_{c}(x) = V_{0,c} e^{\frac{-x^{2}}{\sigma_{c}}}.
\label{eq:pot}
\end{equation}
This means that atomic mirror and the reflecting potential of the atom diode are both implemented with Gaussian potentials $V_{d/m}(x)$.
To avoid having the diode, mirror and imaginary potential all starting in the same point, we assume that all potentials
are at rest until a given time $t_{rest}$ and only then begin moving linearly, i.e. their trajectory is
\begin{eqnarray}
x_{d/m/c} = \left\{ \begin{array}{cc} v_{d/m/c} t_{rest} & 0 \ge t \ge t_{rest}\\
v_{d/m/c} t & t > t_{rest}
\end{array} \right. .
\end{eqnarray}
At final time the velocity-probability distribution is given by $\rho(T,v)= \braket{v|\rho(T)}v$, and the position-probability distribution is given by $\rho(T,x) = \braket{x|\rho(T)}x$.

% ---------------- Figure 7 -----------------------------------------------
\begin{figure}[t]
\begin{center}
(a)\\
\includegraphics[width=0.8\linewidth]{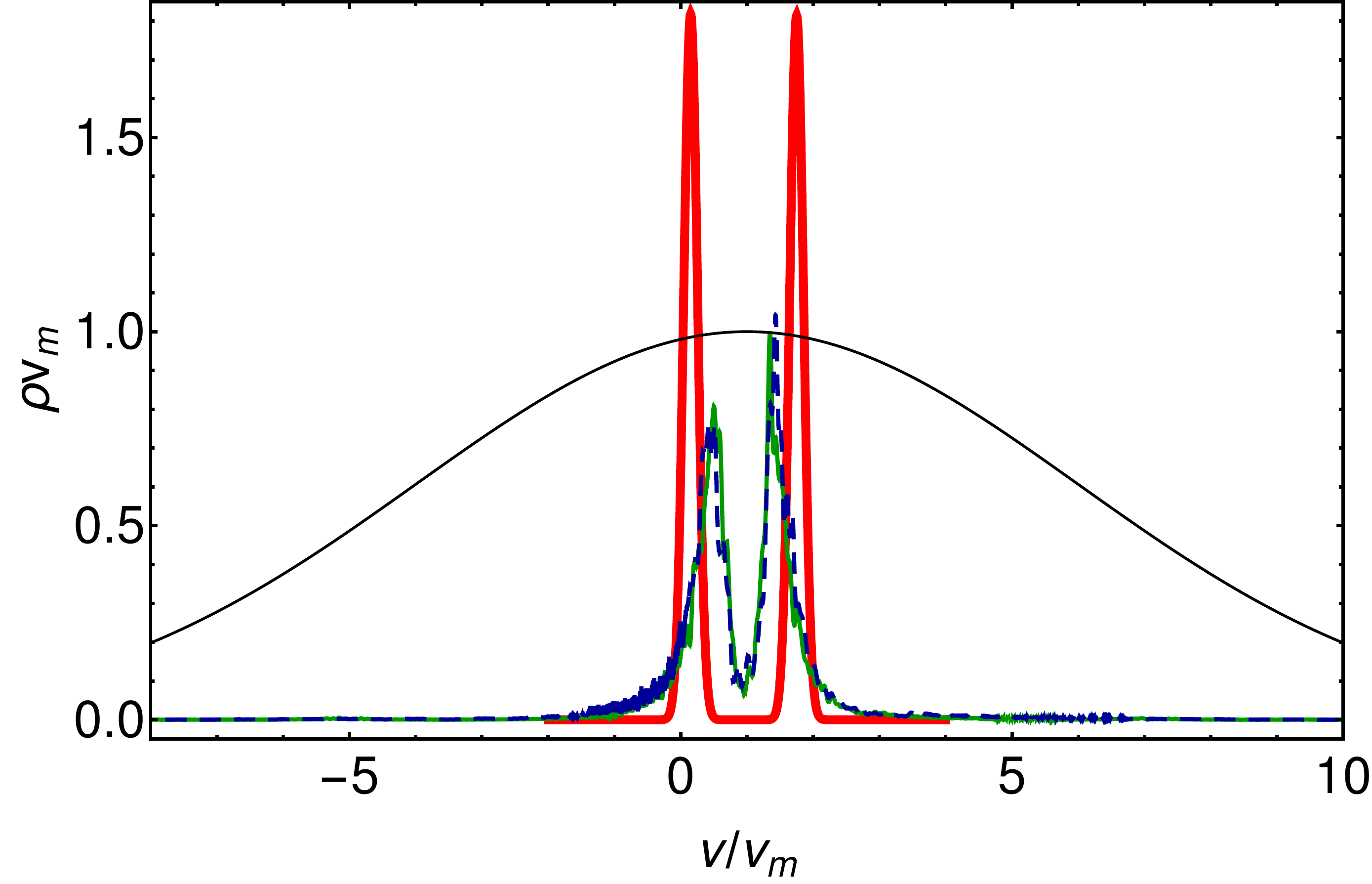}\\
(b)\\
\includegraphics[width=0.8\linewidth]{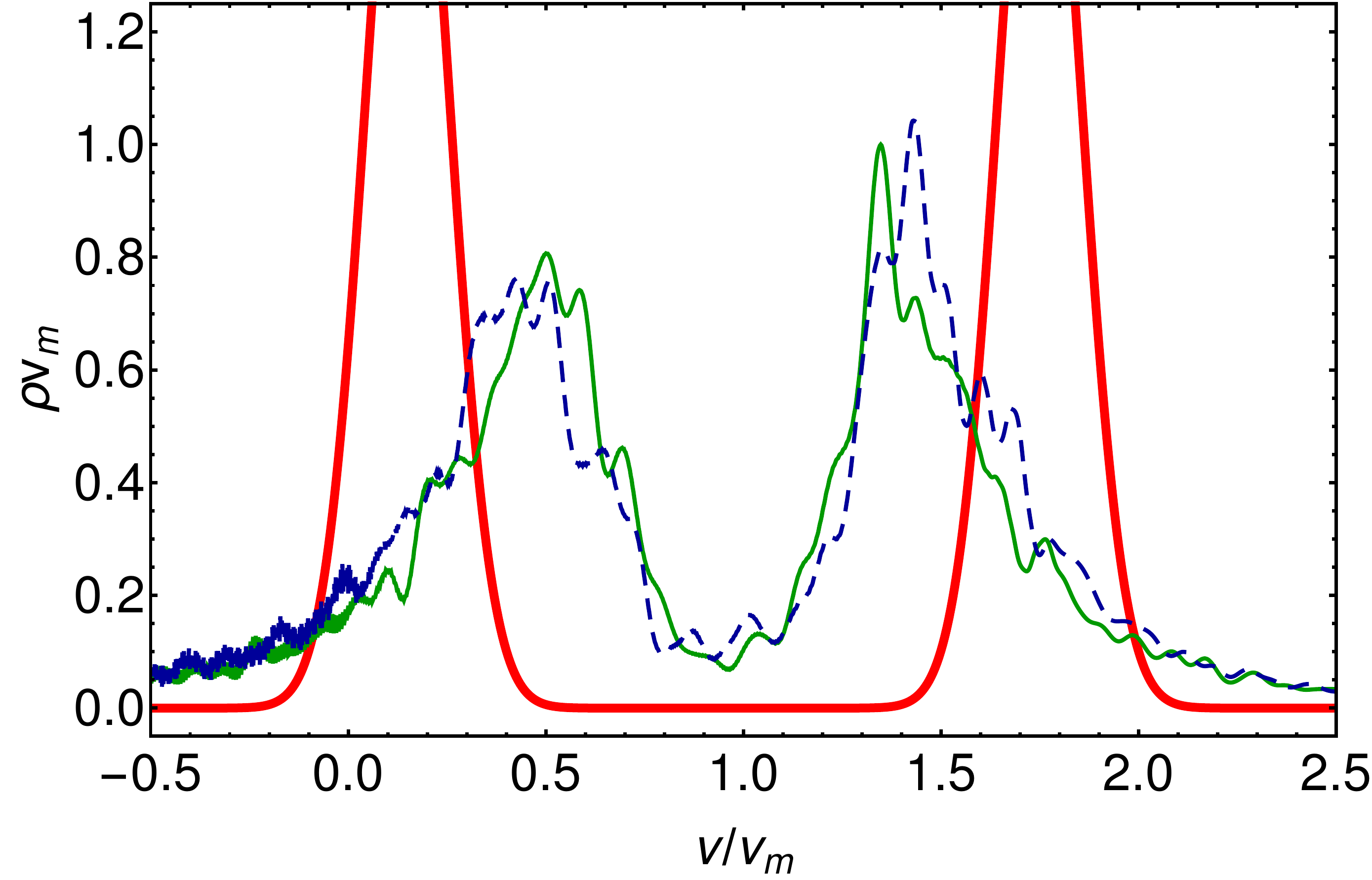}\\
(c)\\
\includegraphics[width=0.8\linewidth]{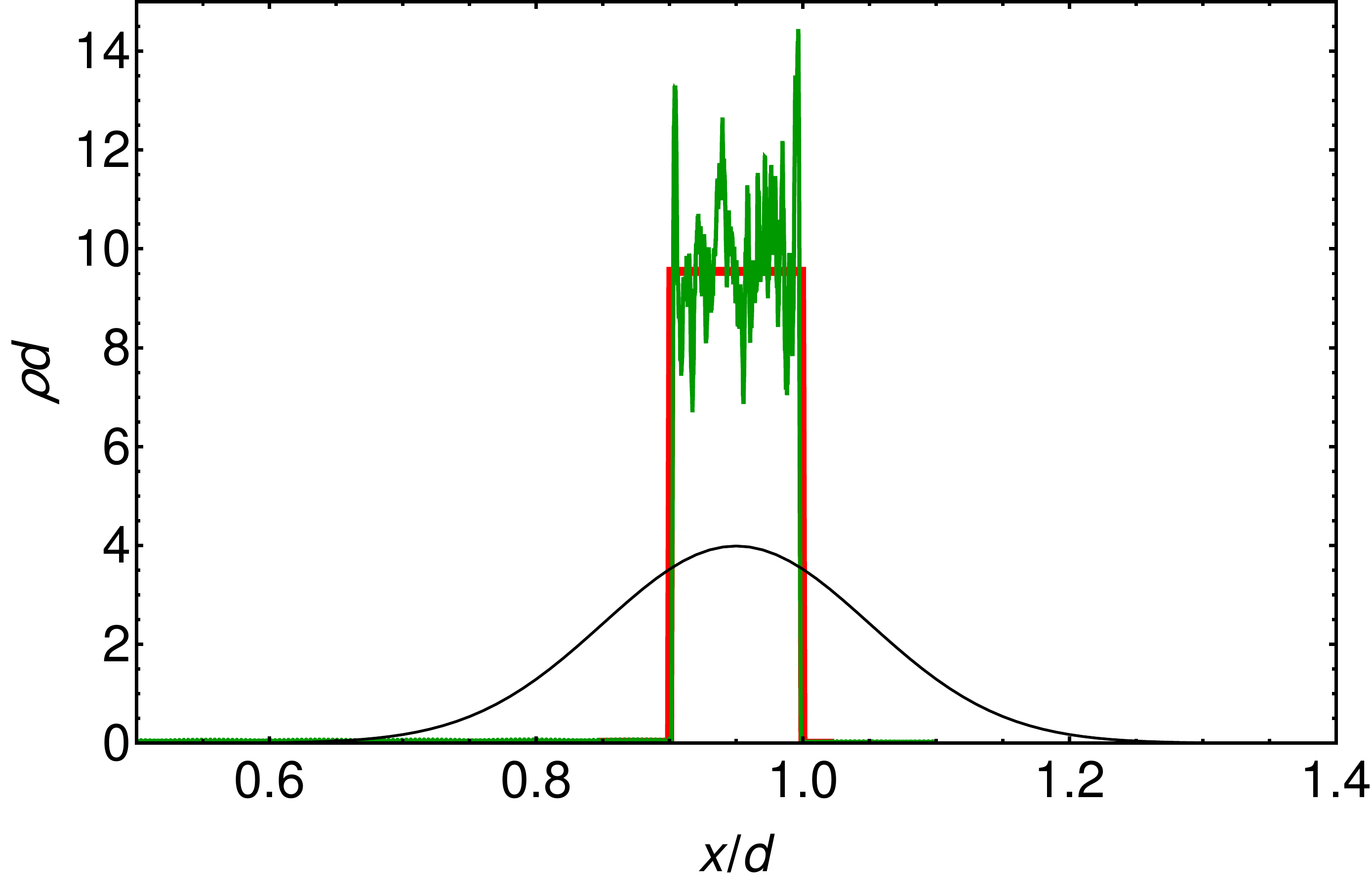}
\end{center}
\caption{Probability distributions: initial distribution (shifted, black, solid lines), final distributions for the classical setting (red, thick line), quantum setting with $v_{0} = 10 v_{m}$ (green, thin line) and quantum setting with $v_{0} = 8 v_{m}$ (blue, dashed line); 
(a) velocity space, (b) velocity space zoomed in and (c) position space. Common parameters: $v_{d} = 0.9 v_m$, $v_0 = 10 v_m$, $\Delta v = 5 v_m$, $x_{0} = -0.8 d$, $\Delta x = 0.1 d$.
Additional parameters in the quantum setting: $V_{0,d/m} = 5 \times 10^{6} \hbar/T$, $V_{0,c} = 4 \times 10^{4}  \hbar/T$, $v_{c} = 0.98 v_m$, 
$\sigma_{c} = 0.0006 d, \sigma_{d} = \sigma_{m} = 0.0001 d$, $m=1000 T\hbar/d^2$.
\label{Figquantum1}}
\end{figure}
% -----------------------------------------------------------------------

% ---------------- Figure 8 ---------------------------------------------
\begin{figure}[t]
\begin{center}
\includegraphics[width=0.8\linewidth]{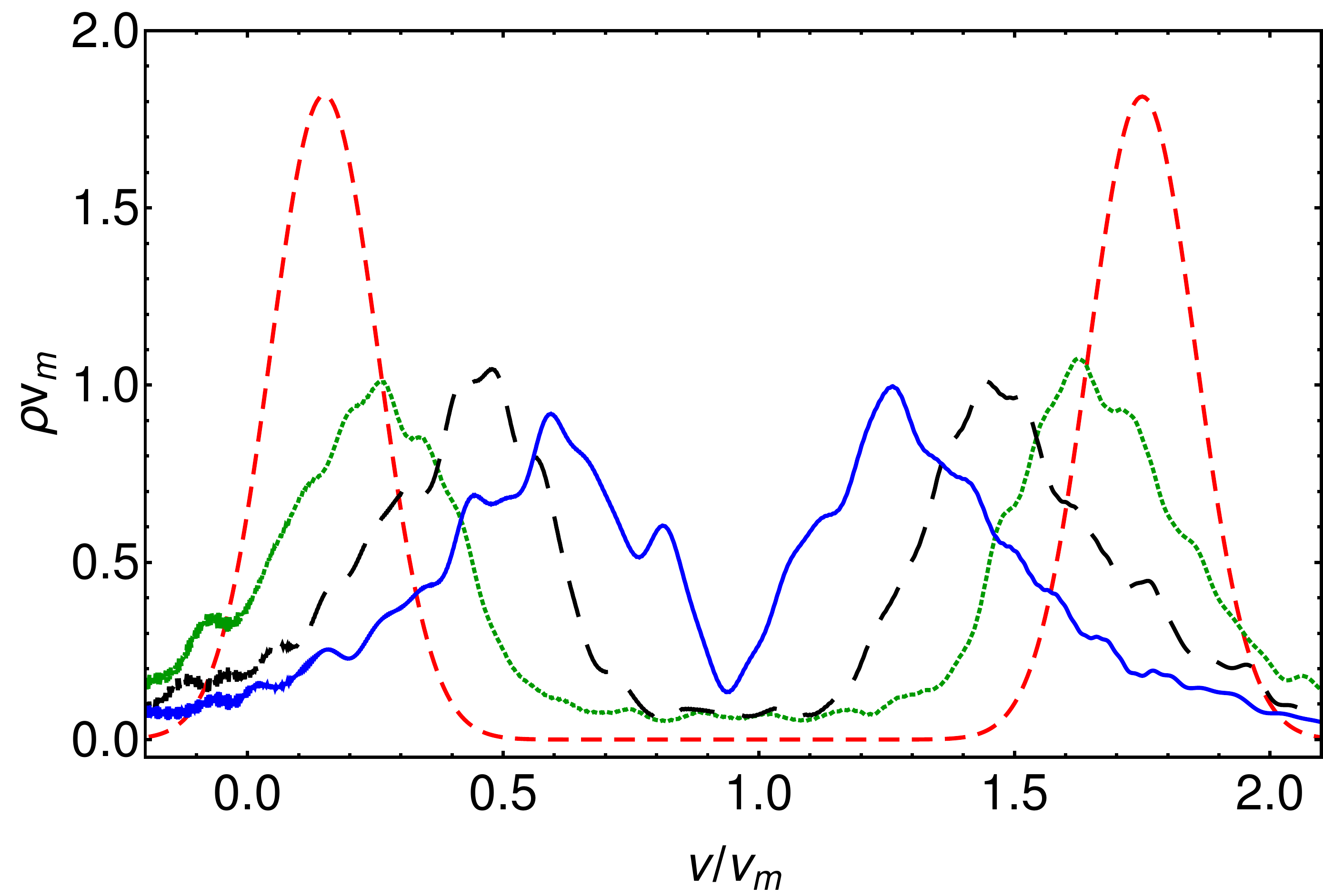} \\
\end{center}
\caption{Final quantum velocity distribution with decreasing $\sigma_{d/m}$: $\sigma_{d/m} = 0.0008  d$ (blue, solid line), $\sigma_{d/m} =0.0004 d$ (black, dotted line), $\sigma_{d/m} =0.0001 d$ (green, solid line) and classical distribution (red, dashed line). $V_{d/m} = 5 \times 10^{5}\hbar/T$, other parameters are the same as in \fref{Figquantum1}. 
\label{decreasing_sigma}}
\end{figure}
% -----------------------------------------------------------------------

\subsection{Results}

In the following, we choose the parameters shown in the caption of \fref{Figquantum1}.
The classical results are independent of the particle mass (as only free motion and ideal, elastic collisions with ideal walls are considered).
The quantum-mechanical result depends on the mass and so we set here $m=1000 T\hbar/d^2$. For example in the case of $^{87}Rb$ if we set $d = 10 \, \mu m$, then $T \approx 0.13 \, m s$ and $v_{0} \approx 0.73 \, m s^{-1}$.

In \fref{Figquantum1} the initial and final velocity distribution are shown and there is a good qualitative correlation between the classical and quantum distributions. In \fref{Figquantum1} (a) we see that in both the quantum and classical distributions are much compressed compared to the original very broad distribution.
As expected the particles are confined between the two walls of the catcher (see \fref{Figquantum1} (c)). Therefore the position distribution is much narrower than the initial distribution, together with the compression in velocity distribution gives us the cooling we desired.
The quantum scheme even retains another interesting property of the classical system; 
we see in \fref{Figquantum1} (b) that, similar to the classical version, the velocity at final time $T$ is almost independent of the initial velocity.

In \fref{Figquantum1} (a) and (b) a difference between the two cases can be seen, the quantum distribution is significantly broader than the classical; further they are less smooth. This appears to be partly because of the quenching of the wave function when it has to transition from being in state $\ket{1}$ to state $\ket{2}$. 

An interesting effect to note however is that the quantum system performs better than the classical. This effect appears to be due to the broadness of our potentials $V_{d/m}$; in the classical simulation we treat these walls as infinitely high but however in the quantum case they have the form of \eref{eq:pot}. 

Heuristically this cooling scheme works through repeated collisions with the mirror/diode and so the effect of the broad potential increases cooling as the particle is reflected  far from the centre of the potential.
Therefore, in \fref{decreasing_sigma}, we examine the effect of reducing $\sigma_{d/m}$. We see that for smaller $\sigma_{d/m}$ we get closer agreement between quantum and classical schemes. This is because for smaller values of $\sigma_{d/m}$ our quantum potentials behave more and more like the infinite potential barriers in the classical case. As there are so many collisions that take place in the diode-mirror system it is quite sensitive to tuning of the parameter $\sigma_{d/m}$, with broader potentials enabling better cooling in the trap.

%---------------------------------------------------------------------------------
%------------------------ Conclusion ---------------------------------------------
%---------------------------------------------------------------------------------

\section{Conclusion}

In this paper we have presented a method for trapping and cooling particles using an atom diode-mirror system. We investigated different trajectories for the diode and the mirror. In particular we found a strong dependence of the efficiency on the trajectory: through classical numerical simulations of linear and square root trajectories we deduced the advantages of the linear scheme for cooling. We propose a way to implement the atom diode and mirror system quantum mechanically; we then applied it to the trapping and cooling of a quantum particle. Through further numerical simulations we demonstrated that we can achieve cooling also in this quantum setting.

\section*{Acknowledgements}
We are grateful to David Rea for useful discussion and commenting on the manuscript.
TD acknowledges support by the Irish Research Council (GOIPG/2015/3195).

\end{document}